\documentclass[reprint,prb,twocolumn]{revtex4-2} 
\usepackage{amsmath}
\usepackage{amsfonts}
\usepackage{graphicx}
\usepackage{color}
\usepackage{xcolor}
\usepackage{float}
\usepackage{soul}
\usepackage{cancel}
\usepackage{hyperref}
\usepackage{dsfont}
\usepackage{graphicx}
\usepackage{blindtext}

\begin{document}
\title {Hybrid light-matter states in topological superconductors coupled to cavity photons}
\author{Olesia Dmytruk$^{1,2}$ and Marco Schir\`o$^{2}$}
\affiliation{$^{1}$ CPHT, CNRS, {{\'E}}cole Polytechnique, Institut Polytechnique de Paris, 91120 Palaiseau, France\\
	$^{2}$ JEIP, UAR 3573 CNRS, Coll\`{e}ge de France, PSL Research University, 11 Place Marcelin Berthelot, 75321 Paris Cedex 05, France}

\date{\today}

\begin{abstract}
We consider a one-dimensional topological superconductor hosting Majorana bound states at its ends coupled to a single mode cavity. In the strong light-matter coupling regime, electronic and photonic degrees of freedom hybridize resulting in the formation of polaritons. We find the polariton spectrum by calculating the cavity photon spectral function of the coupled electron-photon system. In the topological phase the lower in energy polariton modes are formed by the bulk-Majorana transitions coupled to cavity photons and are also sensitive to the Majorana parity. In the trivial phase the lower polariton modes emerge due to the coupling of the bulk-bulk transitions across the gap to photons. Our work demonstrates the formation of polaritons in topological superconductors coupled to photons that contain information on the features of the Majorana bound states.
\end{abstract}

\maketitle

\section{Introduction}

Cavity embedding provides a promising avenue to probe and control quantum materials and devices. On the one hand there is tantalizing possibility of controlling phase transitions and phase diagrams by coupling to a cavity mode, an idea which has received theoretical and experimental attention~\cite{garciavidal2021manipulating,schlawin2022cavity}. Another source of cavity control can arise from the hybridization with finite-frequency modes, leading to new hybrid quasiparticles~--~polaritons~\cite{hopfield1958theory}, which can be then probed and control in novel ways. A wide range of polaritonic modes have been proposed and observed, classified depending on the type of  charged particles in the matter component~\cite{basov2021polariton}. 

A particularly appealing scenario arises when the material has a non-trivial topological character, a feature which can then be enhanced or suppressed~\cite{ciuti2021cavity,appugliese2022breakdown,dmytruk2022controlling,perez2023lightmatter,shaffer2023entanglement} or even generated by the coupling with a cavity and thus transmitted to the emergent polariton excitations~\cite{karzig2015topological}. 
Among topological phases of matter, topological superconductors hosting zero-energy Majorana bound states~\cite{kitaev2001unpaired,alicea2012new,beenakker2013search,prada2020andreev} hold a specially interesting place for their potential for quantum computing~\cite{majorana2015sarma}. The prototype system for topological superconductivity is the Kitaev chain model~\cite{kitaev2001unpaired} describing a one-dimensional $p$-wave superconductor with Majorana bound states emerging at its opposite ends in  the topological phase. Promising platforms for the Majorana bound states are superconductor-semiconductor nanowires~\cite{lutchyn2010majorana,oreg2010helical}, graphene-like systems~\cite{dutreix2014majorana,san2015majorana}, and chains of magnetic atoms~\cite{nadj2013proposal,klinovaja2013topological,braunecker2013interplay}. Signatures of the Majorana bound states in the form of zero-bias peak have been experimentally observed in superconductor-semiconductor nanowire platforms~\cite{mourik2012signatures,deng2012anomalous,das2012zero,churchill2013superconductor,deng2016majorana,de2018electric,PhysRevB.107.245423}. However, theoretical works have demonstrated that the zero-bias peak could arise due to non-Majorana mechanisms~\cite{kells2012near,liu2017andreev,ptok2017controlling,setiawan2017electron,moore2018two,reeg2018zero,vuik2019reproducing,hess2021local,hess2023trivial}.

The idea of using cavities to probe and manipulate the Majorana bound states has been explored in different settings~\cite{cottet2013squeezing,dmytruk2015cavity,dmytruk2016josephson,dartiailh2017direct,cottet2017cavity,trif2018dynamic,trif2019braiding,dmytruk2023microwave,ren2023microwave}. In these cases the cavity plays mainly the role of non-invasive spectroscopic tool to probe the physics of these modes. A different scenario arises potentially in the strong or ultrastrong light-matter coupling regime where polariton modes are formed, which in the case of a topological superconductor could take the form of the Majorana polaritons~\cite{trif2012resonantly,bacciconi2023topological}. 

In this work we study the hybrid light-matter states that emerge by coupling topological superconductors to a single mode cavity. We consider two models of topological superconductors hosting the Majorana bound states: a prototype Kitaev chain model~\cite{kitaev2001unpaired} and a more realistic nanowire model~\cite{lutchyn2010majorana,oreg2010helical}. Hybridization between electronic and photonic states results in formation of polaritons. We focus specifically on the signatures of these polaritonic modes which emerge in the cavity photon spectral function~\cite{mazza2019superradiant,amelio2021optical,dmytruk2021gauge,dmytruk2022controlling,vlasiuk2023cavity}, which is directly measurable in a transmission/reflection experiment~\cite{dmytruk2015cavity,perez2022topology}. We find that the polariton spectrum  is sensitive to the Majorana parity in the topological phase. Moreover, the energies of the polariton modes are different in the trivial and topological phases that could be used to probe the emergence of zero modes in topological superconductor.

The paper is organized as follows. In Sec.~\ref{CouplingTopoLight} we introduce two tight-binding models for topological superconductors and  derive how to couple them to a single mode cavity. Then, in Sec.~\ref{PolaritonSpectrum} we calculate the polariton spectrum of the coupled electron-photon system.  Finally, Sec.~\ref{Conclusions} is devoted to conclusions.

\section{Coupling topological superconductors to light} \label{CouplingTopoLight}

We start by discussing how to couple topological superconductors described by a tight-binding model to a single mode cavity. We consider two  models for topological superconductors: (1) a prototype Kitaev chain~\cite{kitaev2001unpaired}  and (2) an experimentally relevant  nanowire with spin-orbit interaction and proximity-induced superconductivity subject to magnetic field~\cite{lutchyn2010majorana,oreg2010helical}.
Contrary to previously studied tight-binding models for non-superconducting systems~\cite{dmytruk2021gauge,dmytruk2022controlling}, the Kitaev chain (nanowire) models  contain $p$-wave ($s$-wave) superconducting pairing term that pairs two neighboring sites (opposite spins) in the chain.

\subsection{Kitaev chain coupled to cavity}
The Hamiltonian for the Kitaev chain reads~\cite{kitaev2001unpaired},
\begin{align}
	H_K &= -\mu\sum_{j=1}^{N}c_j^\dag c_j - t\sum_{j=1}^{N-1}\left(c_{j}^\dag c_{j+1} + \text{h.c.}\right)\notag\\
	 &+\Delta\sum_{j=1}^{N-1}\left(c_j c_{j+1} + \text{h.c.}\right),
	\label{eq:KitaevH}
\end{align}
where $c_j^\dag$ ($c_j$) are fermionic creation (annihilation) operators at site $j$,  $N$ is the total number of sites in the chain, $\mu$ is the chemical potential, $t$ is the hopping amplitude, and $\Delta$ is a $p$-wave superconducting pairing potential. The Kitaev chain is in the topological (trivial) phase if $|\mu|< 2t$ ($|\mu|> 2t$) 
hosting two Majorana bound states describes by the operators $\gamma_{L(R)} = \gamma_{L(R)}^\dag$. These two Majorana operators form a full fermionic state with $c_M = (\gamma_L - i\gamma_R)/2$ that gives raise to the Majorana occupation $n_M = \langle c_M^\dag c_M\rangle $ that determines its parity. The Majorana occupation $n_M$ can be $0$ or $1$ corresponding  to the even (odd) parity.

Next, we couple the Kitaev chain to a single mode cavity given by the  Hamiltonian $H_{ph} = \omega_c \left(a^\dag a +1/2\right)$, where $a^\dag$ ($a$) is the photonic creation (annihilation) operator and $\omega_c$ is the cavity frequency. The Kitaev chain Hamiltonian $H_{K}$ is coupled to the electromagnetic field described by a homogeneous photonic vector potential $\mathbf{A} = \mathbf{u}_x \left(g/e \right) \left(a+a^\dag\right)$ via the Peierls substitution, which is equivalent to  applying a unitary transformation $U$ to the electronic Hamiltonian~\eqref{eq:NanowireH0} only~\cite{dmytruk2021gauge,dmytruk2022controlling}, $H_{K-ph} = H_{ph} + U^\dag H_{K} U$,
with 
\begin{align}
	U = e^{ i \frac{g}{\sqrt{N}} (a+a^\dag)\sum_{j} R_j c^\dag_{j}c_{j}}.
	\label{eq:UnitaryTransformUK}
\end{align}
Here,  $R_j = j - l_0$, where $l_0 = N/2$ for even $N$.  
Using that
\begin{align}
	&U^\dag c_{m} U = e^{i \frac{g}{\sqrt{N}} (a+a^\dag)R_m} c_{m}, \notag\\
	\label{eq:TransformK}
\end{align}
we find that the superconducting pairing term acquires a site-dependent phase and the full light-matter Hamiltonian reads
\begin{align}
&H_{K-ph} = -\mu\sum_{j=1}^{N}c_j^\dag c_j - \sum_{j=1}^{N-1}\left(t e^{i\frac{g}{\sqrt{N}}(a+a^\dag)} c_j^\dag c_{j+1} + \text{h.c.}\right)\notag\\
&+\sum_{j=1}^{N-1}\left(\Delta e^{i\frac{g}{\sqrt{N}}(2 R_j+1) (a+a^\dag)}  c_j c_{j+1}+  \text{h.c.}\right) \notag\\
&+\omega_c\left(a^\dag a + \dfrac{1}{2}\right).
\label{eq:KitaevCoupled}
\end{align}
Moreover, we note that coupling the superconducting pairing term  to light is equivalent to dressing  $\Delta$ with a phase, $\Delta \rightarrow \Delta e^{i\varphi}$~\cite{pientka2013signatures,cottet2015electron,dmytruk2016josephson}.
The phase $\varphi$ could be found under the assumption that
the $p$-wave pairing term in $H_K$ is inherited from the bulk $s$-wave superconductor underneath the wire. 
In this case, we consider that the instantaneous supercurrent flowing through the bulk superconductor vanishes, 
\begin{equation}
	J_s=\frac{2e}{m}|\psi|^2(\nabla\varphi-2e\mathbf{A})\equiv0\,.
\end{equation}
Here, $m$, $|\psi|^2$, and $\varphi$ are the electronic mass, the density of superconducting electrons in the $s$-wave superconductor, and  its phase, respectively.   The solution of the differential equation  $\nabla\varphi = 2e \mathbf{A}$ gives us $\varphi_j =  2g (a+a^\dag)(j - l_0+1/2)\sqrt{N}$. Here, $\varphi_j$ is chosen such that $\varphi_1 = -\varphi_N$~\cite{perez2022topology}. We note that these two approaches result in the same light-matter Hamiltonian given by Eq.~\eqref{eq:KitaevCoupled}.  Alternatively, light-matter coupling could be included in the problem by starting with a semiconducting nanowire tunnel coupled to a bulk $s$-wave superconductor and assuming that the tunneling hopping is dressed with the Peierls phase~\cite{dmytruk2015cavity}.

\subsection{Superconductor-semiconductor nanowire coupled to cavity} 
We now consider a more realistic model of topological superconductor coupled to photonic cavity. 
The tight-binding Hamiltonian composed of $N$ sites that describes a semiconducting nanowire with Rashba spin-orbit interaction and proximity-induced superconductivity subject to magnetic field reads~\cite{dmytruk2018suppression}
\begin{align}
H_{nw}&=  \sum_{j,\sigma,\sigma'} \Big[
c^\dag_{j+1,\sigma}\left(-t \delta_{\sigma\sigma'}  + i \alpha \sigma^y_{\sigma\sigma'}\right)c_{j,\sigma'}+\Delta c^\dag_{j,\uparrow}c^\dag_{j,\downarrow}
\notag\\
&+\dfrac{1}{2}c^\dag_{j,\sigma}\left[\left(2t - \mu \right) \delta_{\sigma\sigma'}
+ V_Z \sigma^x_{\sigma\sigma'}\right]c_{j,\sigma'} +  {\rm h. c.}\Big],
\label{eq:NanowireH0}
\end{align}
where $c_{j,\sigma}^\dag (c_{j,\sigma})$ is the creation (annihilation) operator acting on electrons with spin $\sigma$ located at  site $j$, $\sigma_{x(y)}$ is the $x$ ($y$) Pauli matrix acting in the spin space, and $t = \hbar^2/\left(2 m^* a_l^2\right)$ is the hopping amplitude, with $m^*$ the effective mass and $a_l$  lattice constant. Here, $\alpha$ is the spin-orbit coupling,  $\Delta$ is the proximity-induced superconducting pairing potential, $\mu$ is the chemical potential, and $V_Z = g^*\mu_B B/2$ is the Zeeman energy, 
with $g^*$ the $g$-factor of the nanowire and $\mu_B$ the Bohr magneton. The nanowire hosts Majorana bound states emerging at the opposite ends of the one-dimensional system if $V_Z > \sqrt{\Delta^2 + \mu^2}$~\cite{lutchyn2010majorana,oreg2010helical}.

Similarly to the Kitaev chain, the light-matter Hamiltonian for the nanowire coupled to a single mode cavity could be obtained by performing the unitary transformation 
$H_{nw-ph} = H_{ph} + U^\dag H_{nw} U$,
with 
\begin{align}
U = e^{ i \frac{g}{\sqrt{N}} (a+a^\dag)\sum_{j\sigma} \chi_j c^\dag_{j\sigma}c_{j\sigma}}.
\label{eq:UnitaryTransformU}
\end{align}
Here,  $\chi_j = j - j_0$  is chosen such that $\chi_1 = -\chi_N$~\cite{perez2022topology}, with $j_0 = \left(N+1\right)/2$ for even $N$.  
Using that
\begin{align}
&U^\dag c_{m\sigma'} U = e^{i \frac{g}{\sqrt{N}} (a+a^\dag)\chi_m} c_{m\sigma'}, \notag\\
\label{eq:Transform}
\end{align}
we find that total light-matter coupling Hamiltonian becomes
\begin{align}
&H_{nw-ph}=  \sum_{j,\sigma,\sigma'} \Big[
c^\dag_{j+1,\sigma}\Big(-t e^{-i\frac{g}{\sqrt{N}}(a+a^\dag)} \delta_{\sigma\sigma'}\notag\\
&+ i \alpha e^{-i\frac{g}{\sqrt{N}}(a+a^\dag)}  \sigma^y_{\sigma\sigma'}  \Big)c_{j,\sigma'}
+\Delta e^{-2i\frac{g}{\sqrt{N}} \chi_j (a+a^\dag)} c^\dag_{j,\uparrow}c^\dag_{j,\downarrow}
\notag\\
&+\dfrac{1}{2}c^\dag_{j,\sigma}\left[\left(2t - \mu \right) \delta_{\sigma\sigma'}
+ V_Z \sigma^x_{\sigma\sigma'}\right]c_{j,\sigma'} +  {\rm h. c.}\Big] \notag\\
&+\omega_c\left(a^\dag a + \dfrac{1}{2}\right).
\label{eq:NanowireCoupled}
\end{align}
In the next section we will discuss the cavity photon spectral function for the two models in Eqs.~\eqref{eq:KitaevCoupled} and \eqref{eq:NanowireCoupled} and highlight the emergence of polariton excitations and their topological signatures.

\section{Polariton spectrum}\label{PolaritonSpectrum}
 In the strong light-matter coupling regime, the electronic and photonic states hybridize giving raise to the formation of new hybrid quasiparticles~--~polaritons.  
 The polariton spectrum can be obtained by computing the cavity photon spectral function
\begin{equation}
A(\omega)=-\dfrac{1}{\pi}\text{Im}\int dt e^{-i\omega t}\left(-i\theta(t)\right)\langle\left[a(t),a^\dag\right]\rangle\,.
\end{equation}
To compute this quantity we follow Refs.~\cite{mazza2019superradiant, dmytruk2021gauge,amelio2021optical,dmytruk2022controlling}, write down the action for the electron-photon problem which we evaluate at the saddle point plus Gaussian fluctuations in the cavity field. Due to gauge-invariance the photon remains incoherent in presence of a uniform vector potential~\cite{nataf2019rashba,andolina2019cavity,guerci2020superradiant,andolina2020theory,dmytruk2021gauge}. The light-matter coupling however gives rise to a self-energy correction for the cavity mode arising from current-current fluctuations of the electronic system. As a result the cavity spectral function takes the form~\cite{ dmytruk2021gauge,dmytruk2022controlling}
\begin{align}
	A(\omega)= -\dfrac{1}{\pi}\dfrac{\chi''(\omega) (\omega +\omega_c)^2}{(\omega^2-\omega_c^2 -2 \omega_c\chi'(\omega))^2 + (2\omega_c \chi''(\omega) )^2},
	\label{eq:Aomegaapp}
\end{align}
where $\chi(\omega) = K(\omega) - \langle J_d \rangle$ is the current-current correlation function, with
\begin{align}
	K(t - t') = -i\theta(t - t')\langle \left[J_p(t), J_p(t')\right]\rangle.
 \label{eq:ParamagneticCorrelation}
\end{align}
Here, $J_p$ ($J_d$) are paramagnetic (diamagnetic) current operators that could be defined from the second-order expansion in $g$~\cite{dmytruk2021gauge,dmytruk2022controlling}
\begin{align}\label{eq:H_Kwph}
H_{K(nw)-ph} \approx \omega_c a^\dag a + H_{K(nw)} + \left(a+a^\dag\right) J_p -  \dfrac{\left(a+a^\dag\right)^2}{2} J_d
\end{align}
and $\theta(t-t')$ is the Heaviside step function.

Polariton spectrum is approximately given by the solutions of the equation~\cite{guerci2020superradiant,dmytruk2021gauge}
\begin{align}
\omega^2\approx \omega_c^2 + 2 \omega_c\chi'(\omega).
\label{eq:PolaritonEnergy}
\end{align}
For $g = 0$ the topological superconductor and cavity photons are fully decoupled and there is a single solution of Eq.~\eqref{eq:PolaritonEnergy} given by $\omega = \omega_c$. 
For finite light-matter coupling $g\neq 0$ electrons and photons are coupled resulting in multiple solutions  that depend both on cavity frequency $\omega_c$ and parameters of the electronic system through the real part of the current-current correlation function $\chi'(\omega)$. Therefore, the resulting polariton energies are sensitive to the properties of the topological superconductor.

We start by deriving the general expression for the current-current correlation function $\chi(\omega)$. Coupling between topological superconductor and cavity photons induces transitions between the Majorana and bulk states in the chain~\cite{dmytruk2015cavity,dmytruk2023microwave,ren2023microwave}. 
These Majorana-bulk transitions could be directly seen as peaks in the imaginary part of the correlation function $K(\omega)$ Eq.~\eqref{eq:ParamagneticCorrelation}. To evaluate $K(\omega)$, we rewrite the fermionic operators $c_{j}$ ($c_{j}^\dag$) in terms of the annihilation (creation) operators $\tilde{c}_n$ ($\tilde{c}_n^\dag$) for the Bogoliubov quasiparticles~\cite{dmytruk2015cavity,dmytruk2023microwave}
\begin{align}
    c_{j} = \sum_{n}\left(u_{j,n}\tilde{c}_n + v_{j,n}\tilde{c}_n^\dag\right),
\end{align}
so that the electronic Hamiltonian \eqref{eq:KitaevH} \eqref{eq:NanowireH0} becomes diagonal $\tilde{H}_{el} = \sum_{n}\epsilon_n \left(\tilde{c}_n^\dag\tilde{c}_n-1/2\right)$. Here, $u_{j,n}$ ($v_{j,n}$) are the electron (hole) components of the eigenvectors and $\epsilon_n$ are the corresponding eigenvalues of the  electronic Hamiltonian, with $n = 1...N$ for the Kitaev chain Hamiltonian \eqref{eq:KitaevH} and $n = 1...2N$ for the superconductor-semiconductor nanowire Hamiltonian \eqref{eq:NanowireH0}. To calculate the expectation value of the diamagnetic current operator $\langle J_d \rangle$ over a bare electronic Hamiltonian \eqref{eq:KitaevH} \eqref{eq:NanowireH0} we rewrite $J_d$ in terms of $\tilde{c}_n$ ($\tilde{c}_n^\dag$) operators and use that $\langle \tilde{c}_{n}^\dag \tilde{c}_m\rangle = f(\epsilon_n)\delta_{n,m}$, with $f(\epsilon_m)$ being the Fermi distribution function. Assuming zero temperature, $f(\epsilon_m)$ reduces to the occupation number $n_m$ that can take values $0$ or $1$ for empty or occupied state. Under this assumption, we arrive at the following expression
\begin{align}
    \langle J_d \rangle = \sum_m j^d_m n_m,
    \label{eq:Jdexpression}
\end{align}
where $j^d_{m} $ is the diagonal matrix element for the diamagnetic current operator between eigenstates corresponding to the eigenvalues $\epsilon_m$.   Defining the Fourier transformation as $K(\omega) = \int e^{i\omega t} K(t)$ and using that $\tilde{c}_m(t) = \tilde{c}_m(0)e^{-i\epsilon_mt}$, we find the general expression for the paramagnetic current correlation function at zero temperature
\begin{align}
    K(\omega) = \sum_{l,m}|j^p_{l,m}|^2\dfrac{n_l-n_m}{\omega+\epsilon_l - \epsilon_m+i\eta}.
    \label{eq:KomegaGeneral}
\end{align}

\begin{widetext}
    
\begin{figure}[t]
\includegraphics[width=\linewidth]{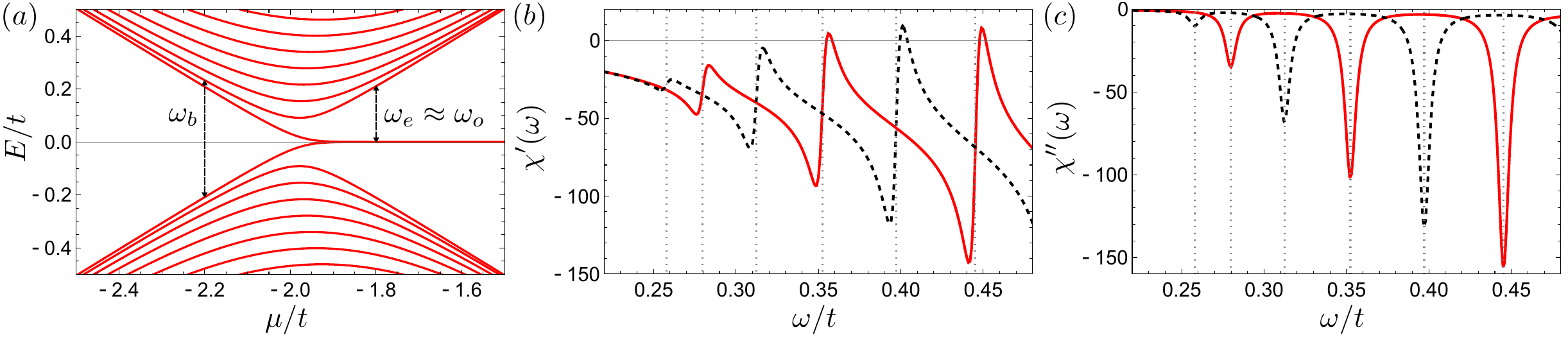}
\caption{(a) Energy spectrum of the Kitaev chain \ref{eq:KitaevH} as a function of the chemical potential $\mu/t$ (red solid lines). Vertical dashed lines indicate the transition frequencies $\omega_b$ and $\omega_{e(o)}$. For the chosen set of parameters Majorana energy $\epsilon_M = 7.44 \times 10^{-7}$ and $\omega_{e} \approx \omega_{o}$. (b) Real part of the current-current correlation function $\chi'(\omega)$ as function of frequency $\omega/t$. Red solid (black dashed) lines correspond to even (odd) Majorana parity. The transition frequencies $\omega_{e}$ are shown in gray vertical dotted lines. (c) Imaginary part of the current-current correlation function $\chi''(\omega)$ as function of frequency $\omega/t$ for even (odd) Majorana parity shown in red solid (black dashed) lines. Gray  vertical dotted lines correspond to the transition frequencies and indicate the position of the peaks.
		Parameters are chosen as $N=100$, $\Delta/t = 1$, $\mu/t = -1.75$ (except in panel (a)), $\eta = 4\times 10^{-3}$.}
  \label{fig:KitaevCurrentCorrelationF}
\end{figure}

\end{widetext}

Here, $j^p_{l,m}$ are the matrix elements of the paramagnetic current and $\eta>0$ is the linewidth of the energy levels.
At zero temperature only bulk states with negative energies are occupied, while $n_l \equiv n_M = 0,1$ for the Majorana states. We note that $K(\omega) = 0$ for $l = m$ making it fully off-diagonal in contrast to $\langle J_d \rangle $.  When the system is in  the topological phase the paramagnetic current correlation function given by Eq.~\eqref{eq:KomegaGeneral} can be rewritten as a sum of three contributions $K(\omega) = K_{BB}(\omega)+K_{BM}(\omega)+K_{MM}(\omega)$, corresponding respectively to transitions between bulk states only ($K_{BB}$), between Majorana and bulk states ($K_{BM}$) and between Majorana states only ($K_{MM}$).
We note that $K_{MM}(\omega) = 0$ since Majorana parity remains conserved in the presence of coupling to photons~\cite{dmytruk2015cavity}. The bulk only contribution in the topological phase (or the total paramagnetic current correlation function in the trivial phase) could be further simplified to
\begin{align}
    &K_{BB}(\omega) = \sum_{\epsilon_{l\neq m} >0}\left(\dfrac{1}{\omega - \omega_{b} +i\eta}- \dfrac{1}{\omega+ \omega_{b} +i\eta}\right)\notag\\
    &\times |j^p_{l,-m}|^2,
    \label{eq:KBB}
\end{align}
where $\omega_{b} = \epsilon_l +\epsilon_m$ is the transition frequency between the bulk states $l$ and $m$, and $j^p_{l,-m}$ is the matrix element between the bulk states with energies $\epsilon_l$ and $-\epsilon_m$. The peaks in the imaginary part of the bulk contribution appear at transition frequencies $\omega_b > 2\Delta_g$, where $\Delta_g$ is the the effective gap in the electronic energy spectrum.

Furthermore, the bulk-Majorana transitions are included in $K_{BM}(\omega)$ term given by
\begin{align}
    &K_{BM}(\omega) = \sum_{\epsilon_l >0}\left(\dfrac{1}{\omega - \omega_{e(0)} +i\eta}- \dfrac{1}{\omega+ \omega_{e(0)} +i\eta}\right)\notag\\
    &\times\left[|j^p_{l,o}|^2(n_M-n_l) + |j^p_{l,e}|^2(1 - n_l-n_M)\right],
    \label{eq:KBM}
\end{align}
where $\omega_{e(0)} = \epsilon_l \pm \epsilon_M$ is the transition frequency between bulk state with occupation number $n_l = 0$ and Majorana state with occupation number $n_M = 0 (1)$  corresponding to even (odd) parity, and $j^p_{l,e(o)}$ is the matrix element between bulk state $l$ and even $e$ (odd $o$) parity Majorana state. The imaginary part of the paramagnetic current correlation function $K''_{BM}(\omega)$ calculated for even parity with $n_M = 0$ has multiple peaks at frequency $\omega_{e}>\Delta_g$ with the amplitude given by $|j^p_{l,e}|^2$, while for $n_M=1$ the peaks are at $\omega_{o}$ with the amplitude given by $|j^p_{l,o}|^2$. Moreover, even in the absence of the overlap between two Majorana bound states $\epsilon_M\approx 0$ the correlation function $K_{BM}(\omega)$ distinguishes between different Majorana parities through the matrix elements $j^p_{l,e(o)}$~\cite{dmytruk2023microwave}.

In the topological phase the cavity spectral function $A(\omega)$ given by Eq.~\eqref{eq:Aomegaapp} depends on the Majorana parity through the different matrix elements entering in  the current-current correlation function $\chi(\omega)$ and, therefore, polariton spectrum could be used to probe Majorana properties. Comparing Eqs.~\eqref{eq:KBM} and~\eqref{eq:KBB} we note that the lowest-energy peaks in the topological and trivial phases appear at frequencies $\omega_{e(o)} \approx \Delta_g$ and $\omega_{b} \approx 2\Delta_g$, respectively, suggesting that the cavity spectral function could be also used to differentiate between two phases. 
 \subsection{Polaritons in Kitaev chain coupled to photons}
We start discussing the cavity spectral function for the Kitaev chain, Eq.~(\ref{eq:KitaevCoupled}). In this case the paramagnetic and diamagnetic current operators could be found from Eq.~\eqref{eq:KitaevCoupled}:
\begin{align}
	J_p = i \dfrac{g}{\sqrt{N}} \sum_j\Big[- t   c^\dag_{j}c_{j+1} +  2\Delta  (R_j +1/2) c_{j}c_{j+1}- \rm{h.c.}\Big]
	\label{eq:ParamagneticCurrentK}
\end{align}
and 
\begin{align}
	J_d =  \dfrac{g^2}{N} \sum_j\Big[-tc^\dag_jc_{j+1} 
 + 4\Delta  ( R_j+1/2)^2  c_{j}c_{j+1}+ \rm{h.c.}\Big],
	\label{eq:DiamagneticCurrentK}
\end{align}
where we see that in addition to the usual contribution from single particle hopping there is also a term coming from the superconducting pairing. We emphasize that this current is not associated to a conserved charged in the Kitaev model, which only enjoys a discrete $Z_2$ parity symmetry. However, it is the natural object entering the response of the system to the cavity vector potential, see Eq.~(\ref{eq:H_Kwph}).

\begin{figure}[b]
\includegraphics[width=\linewidth]{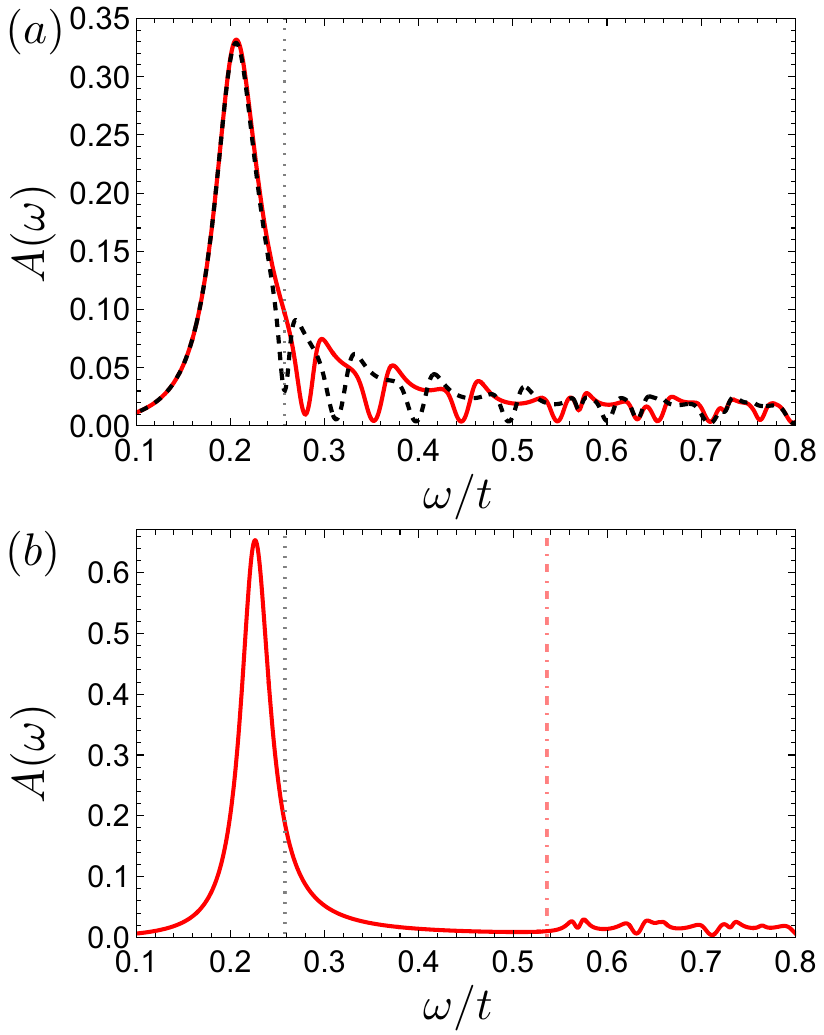}
 \caption{Cavity spectral function $A(\omega)$  as function of frequency $\omega/t$ for $g = 0.1$. (a) In the topological phase ($\mu/t = -1.75$) red solid (black dashed) line corresponds to even (odd) Majorana parity. Vertical gray dotted line indicates the cavity frequency $\omega_c$  fixed to be in resonance with the first bulk-Majorana transition. (b) In the trivial phase ($\mu/t = -2.25$) there is a large peak emerging at $\omega_c$ and smaller peaks appearing at $\omega>2\Delta_g$. Vertical gray dotted line corresponds to $\omega_c$ and pink dotdahsed line indicates the first bulk-bulk transition.
		Other parameters are the same as in Fig.~\ref{fig:KitaevCurrentCorrelationF}.}
  \label{fig:KitaevSpectralFunctionCross}

\end{figure}

To find the cavity spectral function we first calculate the current-current correlation function using Eqs.~\eqref{eq:Jdexpression} and~\eqref{eq:KomegaGeneral}. In Fig.~\ref{fig:KitaevCurrentCorrelationF}~(b) we plot the real part of correlation function $\chi'(\omega)$ as a function of frequency $\omega$. Vertical dotted lines indicate the bulk-Majorana transition frequencies $\omega_{e(o)}$. For the Kitaev chain in the topological phase $\epsilon_M\approx 0$ and therefore $\omega_{e} \approx \omega_{o}$. We find that $\chi'(\omega)$ has different oscillation amplitudes for even and odd Majorana parities stemming from the difference in the matrix elements $j^p_{l,e}$. Next, we numerically evaluate  the imaginary part of the correlation function $\chi''(\omega)$ (see Fig.~\ref{fig:KitaevCurrentCorrelationF}~(c)). The function $\chi''(\omega)$ has multiple peaks at resonant frequencies $\omega_{e(o)}$ that differ for two parities, similarly to the features present in $\chi'(\omega)$.
Therefore, the current-current correlation function $\chi(\omega)$ is a good marker to  distinguish between two Majorana parities in the topological phase.

\begin{figure}[!htb]
  \includegraphics[width=\linewidth]{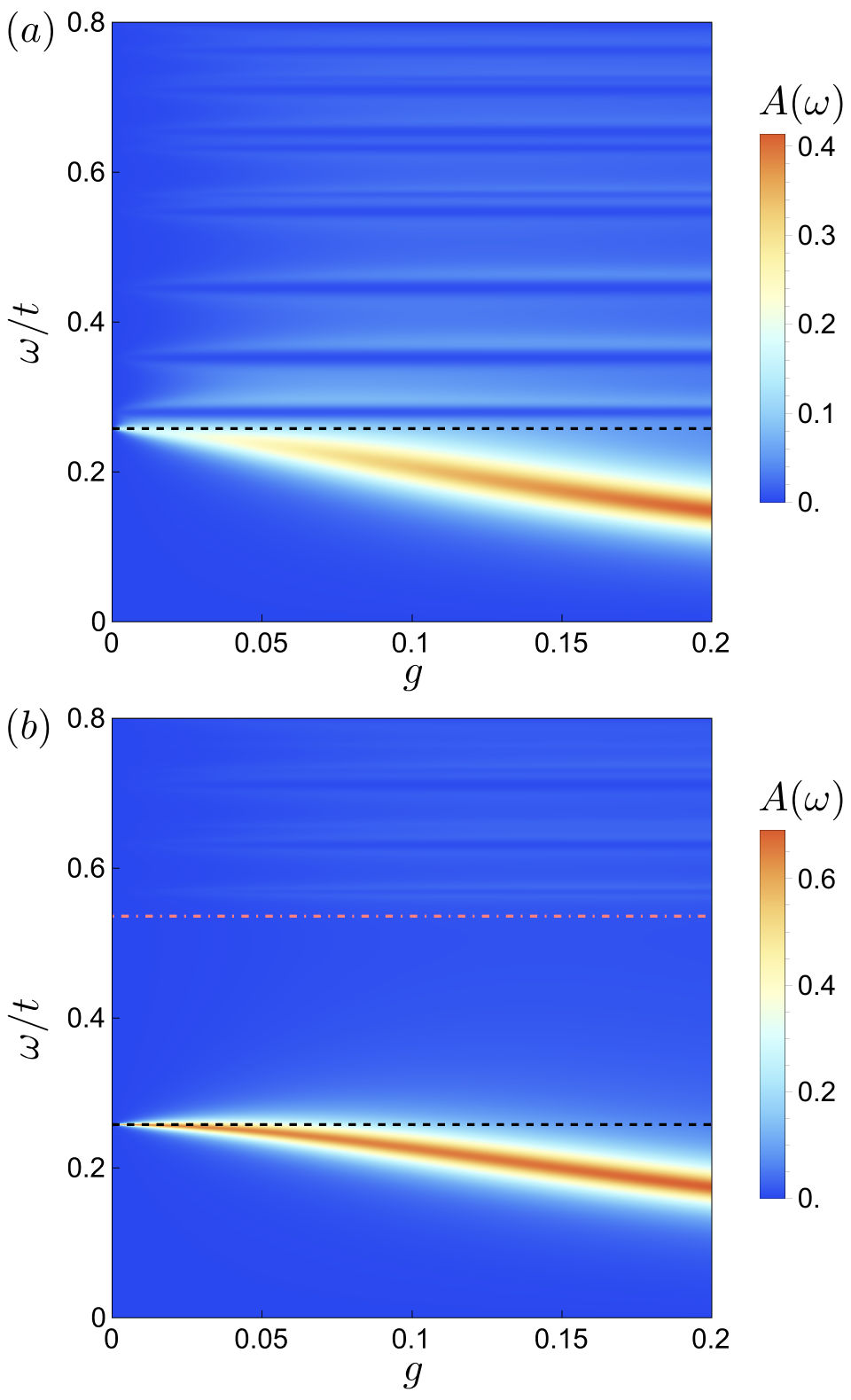}
  \caption{  Spectral function $A(\omega)$ as a function of $g$  and $\omega/t$. Horizontal black dashed line corresponds to frequency $\omega_c$ chosen to be equal to first bulk-Majorana transition frequency and  horizontal pink dotdashed line corresponds to $\omega_b$ in the trivial phase.  (a) In the topological phase with $\mu/t = -1.75$ the lowest polariton branch originating at $\omega = \omega_c$ for $g = 0$ goes down as $g$ is increased. White horizontal lines  corresponding to bulk-Majorana transitions coupled with photons that appear at frequencies $\omega > \Delta_{g}$.
 (b) In the trivial phase with $\mu/t = -2.25$, the lowest polariton branch appears at $\omega = \omega_c$. In contrast to the topological phase, white horizontal lines correspond to bulk-bulk transitions and appear at $\omega> 2\Delta_g$. In two phases white horizontal lines corresponding to bulk-Majorana (a) and bulk-bulk (b) transitions emerge at different frequencies signalling the presence of zero-energy states in the topological phase. Other parameters are the same as in Fig.~\ref{fig:KitaevCurrentCorrelationF}.}
  \label{fig:SpectralFunctionKitaev}

\end{figure}

Given the above results for the current-current correlator we can now focus on the cavity photon spectral function \eqref{eq:Aomegaapp}. We plot $A(\omega)$  as a function of frequency in Fig.~\ref{fig:KitaevSpectralFunctionCross}~(a) at a fixed light-matter coupling $g$ for different parities in topological phase. 
The current-current correlation function is calculated for a finite-length Kitaev chain and has many resonances (see Fig.~\ref{fig:KitaevCurrentCorrelationF}~(b)), therefore, Eq.~\eqref{eq:PolaritonEnergy} has multiple solutions for polariton energies corresponding to peaks in $A(\omega)$. Moreover, the polariton spectrum in topological phase depends on the Majorana parity through $\chi'(\omega)$. The cavity spectral function has different patterns for two parities and can distinguish between the parities. We further compute the cavity spectral function in the trivial phase [see Fig.~\ref{fig:KitaevSpectralFunctionCross}~(b)] for the same light-matter coupling strength $g$ and the effective gap $\Delta_g$. We find that $A(\omega)$ has a sharp peak around the cavity frequency $\omega_c$ as in the topologically nontrivial phase. However, we note that contrary to topological case small peaks emerge at frequencies larger than $2\Delta_g$ corresponding to bulk-bulk transition across the gap in the system.

In Fig.~(\ref{fig:SpectralFunctionKitaev})~(a) we plot the cavity spectral function for the Kitaev chain in  the topological phase as a function of frequency and  light-matter coupling. We consider a cavity frequency in resonance with the first bulk-Majorana transition for the even parity ($\omega_c = \omega_{e}$). We see that for low frequency there is a broad peak which shifts towards lower frequencies upon increasing $g$. At higher frequencies on the other hand we recognize sharp features associated to transitions between Majorana and bulk states. Next, we calculate $A(\omega)$ for the Kitaev chain in the trivial phase [see Fig.~(\ref{fig:SpectralFunctionKitaev})~(b)]. As discussed for the topological phase there is a broad peak that originates at $\omega = \omega_c$ for $g = 0$ and further broaden as the light-matter coupling strength is increased. However, in the trivial phase the current-current correlation function $\chi(\omega)$ that enters Eq.~\eqref{eq:Aomegaapp} has resonances only at frequencies $\omega_{b}>2\Delta_g$. Therefore, other polariton modes appear only at $\omega> 2\Delta_g$.  Comparing the 
cavity spectral function calculated in the topological phases we note the distinct features between the two, namely that the sharp features of the transitions between Majorana (bulk) - bulk states appear at different energy scales of $\Delta_g$ (2$\Delta_g$). 
Therefore, the polariton spectrum could be potentially used as a way to probe zero-energy states in topological superconductors.

\begin{figure}[t!] 
	\includegraphics[width=\linewidth]{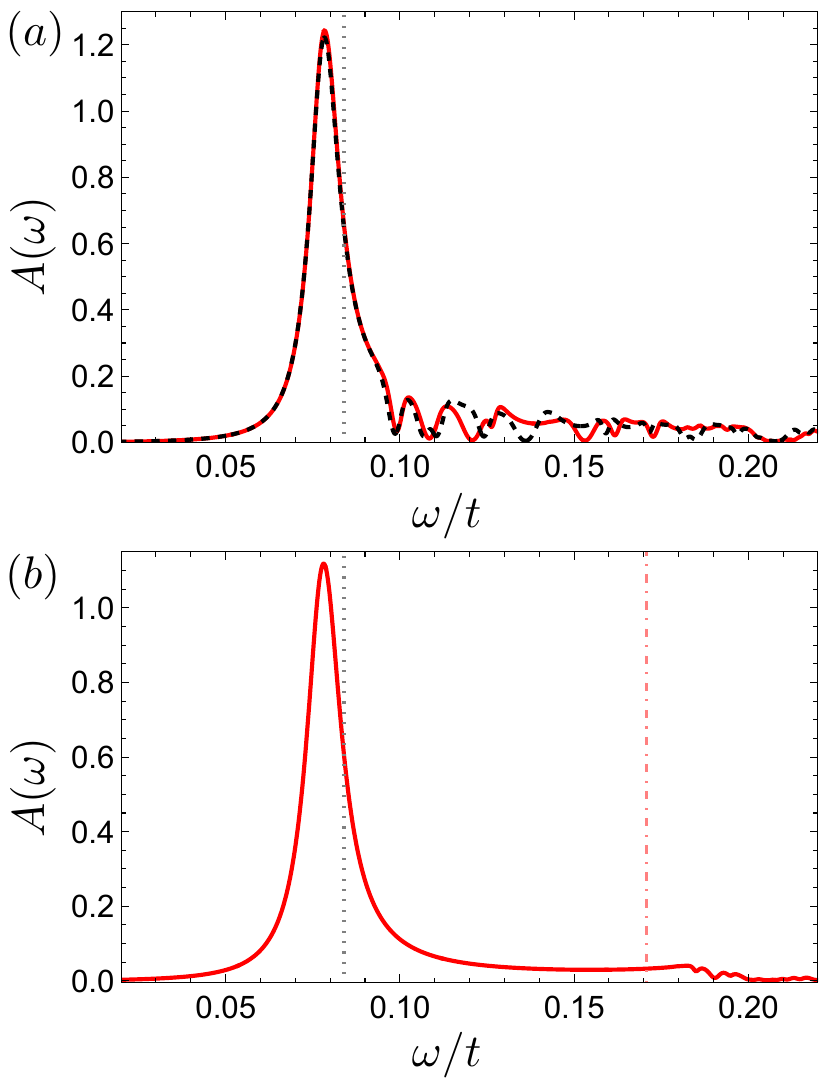}
	\caption{Cavity spectral function $A(\omega)$ of the nanowire  as function of frequency $\omega/t$  for  light-matter coupling strength $g = 0.05$. (a) Red solid (black dashed) lines correspond to $n_M = 0$ ($n_M = 1$) in the topological phase with $V_Z/\Delta = 1.8$. Gray vertical dotted line indicates the cavity frequency $\omega_c$ resonant with the first bulk-Majorana transition at $\omega_e\approx \omega_{o}$ ($\epsilon_M/t = 10^{-6}$). (b) $A(\omega)$ for the nanowire in the trivial phase with $V_Z/\Delta = 0.2$. Vertical gray dotted line indicates $\omega_c$ and pink dotdashed line signals the position of the first bulk-bulk transition frequency $\omega_b$. 
		Other parameters are fixed as $N=100$, $\Delta/t=0.1$,  $\mu = 0$, $V_Z/\Delta = 1.8$, $\alpha/t = 0.4$, and $\eta/t = 10^{-3}$. }
	\label{fig:SpectralFunctionNanowireCross}
\end{figure}

\subsection{Polaritons in nanowire coupled to photons}

We now move to the superconductor-semiconductor nanowire model, Eq.~\eqref{eq:NanowireCoupled}, for which the   paramagnetic and diamagnetic current operators read respectively
\begin{align}
	&J_p = i \dfrac{g}{\sqrt{N}} \sum_j\Big[ t \left(   c^\dag_{j+1\uparrow}c_{j\uparrow} + c^\dag_{j+1\downarrow}c_{j\downarrow} \right) \nonumber\\
	&+  \alpha \left(c^\dag_{j+1\uparrow}c_{j\downarrow} - c^\dag_{j+1\downarrow}c_{j\uparrow}
	\right) - 2 \Delta  \chi_j  c^\dag_{j\uparrow}c^\dag_{j\downarrow}- \rm{h.c.}\Big]
	\label{eq:ParamagneticCurrent}
\end{align}
and 
\begin{align}
	&J_d =  \dfrac{g^2}{N} \sum_j\Big[- t \left(   c^\dag_{j+1\uparrow}c_{j\uparrow} + c^\dag_{j+1\downarrow}c_{j\downarrow} \right) \nonumber\\
	&+  \alpha \left(c^\dag_{j+1\uparrow}c_{j\downarrow} - c^\dag_{j+1\downarrow}c_{j\uparrow}
	\right) + 4 \Delta  \chi_j^2  c^\dag_{j\uparrow}c^\dag_{j\downarrow}+ \rm{h.c.}\Big].
	\label{eq:DiamagneticCurrent}
\end{align}

To find the cavity spectral function of the nanowire model, we proceed in the same way as for the Kitaev chain. The real and imaginary part of $\chi(\omega)$ has similar structure to Fig.~\ref{fig:KitaevCurrentCorrelationF}, but the position and amplitude of the peaks  are less homogeneous due to more involved energy spectrum of the nanowire.

\begin{figure}[t!] 
		\includegraphics[width=\linewidth]{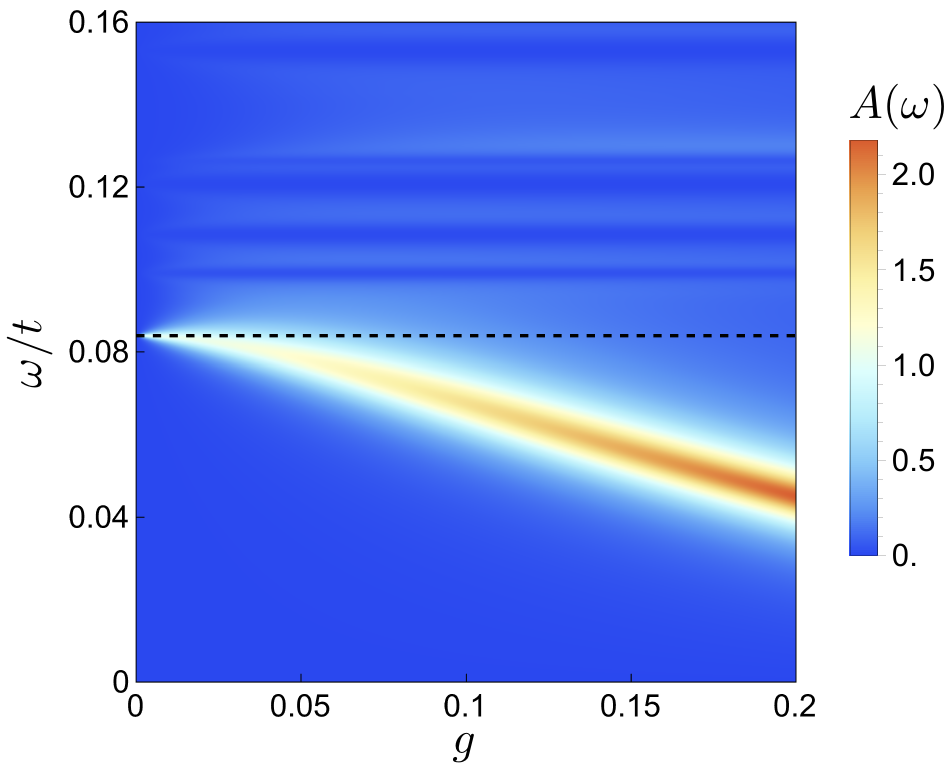}
		\caption{Cavity spectral function $A(\omega)$ of the nanowire as a function of the light-matter coupling $g$ and frequency $\omega/t$. Black dashed line indicates the value of cavity frequency $\omega_c  = \omega_{e}$ (resonant with the first bulk-Majorana transition for even parity). White horizontal lines correspond to bulk-Majorana transitions and emerge at frequencies $\omega>\Delta_g$. Other parameters are same as in Fig.~\ref{fig:SpectralFunctionNanowireCross}.} 
	\label{fig:SpectralFunctionNanowire}
	
\end{figure}

In Fig.~\ref{fig:SpectralFunctionNanowireCross} we plot the cavity spectral function for the nanowire model as a function of frequency $\omega$ for a fixed value of the light-matter coupling $g$. In the topological phase we consider different parities depicted in solid red and black dashed line. The cavity spectral function has a large peak around the cavity frequency $\omega_c$ resonant with the lowest bulk-Majorana transition frequency $\omega_{e}\approx \omega_{o}$ and multiple smaller peaks corresponding to higher in energy bulk-Majorana transitions appearing at $\omega>\Delta_g$  [see Fig.~\ref{fig:SpectralFunctionNanowireCross}~(a)]. 
Considering the superconductor-semiconductor nanowire in the trivial phase coupled to photonic cavity, we find that the cavity spectral function has a sharp peak originating at the frequency $\omega_c$ and multiple smaller peaks at frequencies $\omega>2\Delta_g$ that stem from the bulk-bulk transitions in the nanowire [see Fig.~\ref{fig:SpectralFunctionNanowireCross}~(b)]. Similar features were found for the Kitaev chain [see Fig.~\ref{fig:KitaevSpectralFunctionCross}~(b)] and allow one to probe the presence of zero-energy modes in the topological superconductor.

Finally, we present $A(\omega)$ for the nanowire in the topological phase as a function of frequency and light-matter coupling strength in Fig.~\ref{fig:SpectralFunctionNanowire}. By choosing the cavity frequency to be equal to the first bulk-Majorana transition frequency, we find the appearance of a broad low-frequency polariton mode that goes down in $\omega$ with increasing $g$. Higher-frequency polariton modes appear due to coupling between higher bulk-Majorana transitions and photons showing a dense pattern of modes. Similar behaviour was found for the Kitaev chain (see Fig.~\ref{fig:SpectralFunctionKitaev}).

\section{Conclusions} \label{Conclusions}
In this work, we studied  the topological superconductor coupled to cavity photons. We calculated the cavity spectral function of the  electron-photon system that revealed the polariton spectrum of the hybrid system. The peaks in cavity spectral function appear at different energy scales for the electronic chain in the trivial and topological phase. Moreover, in the topological phase associated with the presence of the Majorana bound states the polariton spectrum has different pattern for two Majorana parities. Therefore, cavity spectral function could be used to probe topological properties of the electronic chain.

\begin{acknowledgements} O.D. acknowledges helpful discussions with Jelena Klinovaja, Daniel Loss, Pascal Simon and Mircea Trif. This project has received funding from the European Union’s Horizon 2020 research and innovation programme under the Marie Skłodowska-Curie grant agreement No 892800.
This project has received funding from the European Research Council (ERC) under the European Union’s Horizon 2020 research and innovation programme (Grant agreement No. 101002955 — CONQUER).
\end{acknowledgements}

\end{document}